\documentclass[%
 reprint,
 amsmath,amssymb,
 nofootinbib
]{revtex4-1}

\usepackage{natbib}
\usepackage[english]{babel}

\usepackage{amsmath,amssymb,amsfonts}
\usepackage{physics}
\usepackage{dsfont}
\usepackage{graphicx}

\usepackage{tikz}
\usepackage{subcaption}

\usetikzlibrary{arrows,shapes,trees}

\newcommand{\ra}{\rightarrow}
\newcommand{\Hom}{{\rm Hom}}

\newcommand{\End}{{\rm End}}

\newcommand{\im}{{\rm im}}

\newcommand{\CC}{{\mathbb C}}
\newcommand{\ZZ}{{\mathbb Z}}

\newcommand{\cP}{{\mathcal P}}

\newcommand{\cA}{\mathcal A}

\newcommand{\cC}{\mathcal C}

\newcommand{\cT}{{\mathcal T}}

\newcommand{\Q}{{\mathcal Q}}

\usepackage[left=0.76in,right=0.76in,top=0.76in,bottom=0.76in]{geometry}

\usepackage{dcolumn}
\usepackage{bm}

\begin{document}

\preprint{APS/123-QED}

\title{Topological Field Theory and Matrix Product States}
\author{Anton Kapustin, Alex Turzillo, Minyoung You \\
{\it California Institute of Technology} \\ {\it 1200 E California Blvd, Pasadena, CA 91125}}

\begin{abstract}

It is believed that most (perhaps all) gapped phases of matter can be described at long distances by Topological Quantum Field Theory (TQFT). On the other hand, it has been rigorously established that in 1+1d ground states of gapped Hamiltonians can be approximated by Matrix Product States (MPS). We show that the state-sum construction of 2d TQFT naturally leads to MPS in their standard form. In the case of systems with a global symmetry $G$, this leads to a classification of gapped phases in 1+1d in terms of Morita-equivalence classes of $G$-equivariant algebras. Non-uniqueness of the MPS representation is traced to the freedom of choosing an algebra in a particular Morita class.  In the case of Short-Range Entangled phases, we recover the group cohomology classification of SPT phases. 

\begin{description}
\item[PACS numbers]
71.27.+a, 02.40.Re
\end{description}

\end{abstract}


\maketitle

\section{Introduction and Overview}

It is a widely held belief that the universal long-distance behavior of a  quantum phase of matter at zero temperature can be encoded into an effective field theory.\footnote{It is hard to make this rigorous since neither the notion of a phase of matter nor that of an effective field theory has been formalized.} In the case of gapped phases of matter, the extreme infrared should be described by a Topological Quantum Field Theory. 

It has been rigorously shown that the ground state of any gapped 1+1d Hamiltonian with a short-range interaction can be approximated by a Matrix Product State.\cite{Hastings} This representation is very efficient, especially in the translationally-invariant case, and is well-suited to the Renormalization Group analysis. In particular, it leads to a classification of Short-Range Entangled Phases of 1+1d matter in terms of group cohomology.
\cite{ChenGuWenone,ChenGuWentwo,FidkowskiKitaev}

It is natural to ask about the connection between these two approaches to gapped phases of matter. In this note we answer this question in the case of bosonic systems with a finite symmetry $G$. For simplicity, we assume that all elements of $G$ act unitarily (i.e. we do not allow for time-reversing symmetries). The case of time-reversing symmetries and fermionic phases will be addressed in separate publications.\cite{KTY2}

In brief, the results are as follows. We show that a standard-form  MPS is naturally associated with a module  $M$ over a finite-dimensional semisimple algebra $A$. The universality class of the MPS depends only on the center $Z(A)$. On the other hand, every unitary 2d TQFT has a state-sum construction which uses a semisimple algebra as an input. Further, given a module $M$ over this algebra, one naturally gets a particular state in the TQFT space of states. We show that this state is precisely the MPS associated to the pair $(A,M)$. Since the TQFT depends only on $Z(A)$, we reproduce the fact that the universality class of the MPS depends only on $Z(A)$. 

In the case of an MPS with a symmetry $G$, a similar story holds. A $G$-equivariant MPS is encoded in a $G$-equivariant module $M$ over a $G$-equivariant semisimple algebra $A$. Such an algebra can be used to give a state-sum construction of a $G$-equivariant TQFT, while every $G$-equivariant module $M$ gives rise to a particular state. This state is an equivariant MPS state. Again, different $A$ can give rise to the same TQFT. This leads to an equivalence relation on $G$-equivariant algebras which is a special case of Morita equivalence. 

An indecomposable phase with symmetry $G$ is therefore associated with a Morita-equivalence class of indecomposable $G$-equivariant algebras. The classification of such algebras is well known\cite{Ostrik} and leads to an (also well-known \cite{ChenGuWenone,ChenGuWentwo,FidkowskiKitaev}) classification of bosonic 1+1d gapped phases of matter with symmetry $G$. In the special case of Short-Range Entangled gapped phases, we recover the group cohomology classification of SPT phases.

\section{Matrix Product States at RG Fixed Points}

\subsection{Matrix Product States}\label{sec:mps}

In this section, we review Matrix Product States (MPS) and extract the algebraic data that characterizes them at fixed points of the Renormalization Group (RG). We find that a fixed point MPS is described by a module over a finite-dimensional  semisimple algebra. We discuss the notion of a gapped phase and argue that they are classified by finite-dimensional semisimple commutative algebras. Given a fixed point MPS and the corresponding semisimple algebra $A$, the commutative algebra characterizing the gapped phase is the center of $A$, denoted $\cA = Z(A)$.

The models we consider are defined on Hilbert spaces that are tensor products of finite-dimensional state spaces $A$ on the sites of a 1D chain. We are interested in Hamiltonians with an energy gap that persists in the thermodynamic limit of an infinite chain. A large class of examples of gapped systems come from local commuting projector (LCP) Hamiltonians; that is, $H=\sum h_{s,s+1}$, where the $h_{s,s+1}$ are projectors that act on sites $s$, $s+1$ and commute with each other. Since the local projectors commute, an eigenstate of $H$ is an eigenstate of each projector. It follows that the gap of $H$ is at least $1$. Thus LCP Hamiltonians are gapped in the thermodynamic limit. In one spatial dimension, ground states of gapped Hamiltonians are efficiently approximated by an ansatz called a matrix product state (MPS),\cite{Hastings} which we recall  below.\footnote{We only consider translationally-invariant MPS.} From each MPS, one can construct a gapped \emph{parent Hamiltonian} that has the MPS as a ground state.\cite{FNW} At RG fixed points, which we consider below, the parent Hamiltonian is an LCP Hamiltonian. To discuss and classify 1D gapped Hamiltonians, it suffices to consider the parent Hamiltonians of the MPS that approximate their ground states.

Consider a closed chain of $N$ sites, each with a copy of a \emph{physical} Hilbert space $A \simeq\CC^d$ and two copies $V^L$, $V^R$ of a \emph{virtual} space $\CC^D$. We identify $V^L=V$ and $V^R=V^*$ and choose a Hilbert space structure on $V$. Between each adjacent pair $(s,s+1)$ of sites, place the maximally entangled state\begin{equation}\ket{\omega}_{s,s+1}=\sum_{i=1}^D\ket{i}\otimes\ket{i}\in V^R_s\otimes V^L_{s+1}\end{equation}
An MPS tensor\footnote{More generally, the tensors $\cP_s$ may depend on the site index $s$. But any translationally-invariant state has an MPS representation with a site-independent tensor.\cite{MPS2}} is a linear map $\cP:V^L\otimes V^R\rightarrow A$. The MPS associated to $\cP$ is the state\begin{align}\ket{\psi_\cP}=&\left(\cP_1\otimes\cP_2\otimes \cdots\otimes\cP_N\right)\nonumber\\&\hspace{5mm}\left(\ket{\omega}_{12}\otimes\ket{\omega}_{23}\otimes\cdots\otimes\ket{\omega}_{N1}\right)\end{align}in $A^{\otimes N}$. Since $\ket{\psi_\cP}$ lies in the image of $\cP^{\otimes N}$, we do not lose generality by truncating $A$ to $\im\, P$. We will assume we have done so in the following. Equivalently, we assume that the adjoint MPS tensor $T=\cP^\dagger$ is injective\footnote{To avoid confusion, we stress that injectivity of $T$ is unrelated to the notion of an injective MPS in the sense of \cite{MPS}. In particular, while we will always assume that $T$ is injective, we will not assume that the the ground state of the parent Hamiltonian is unique.}. The MPS wavefunction can be expressed as a trace of a product of matrices, hence its name. In the basis $\{e_i\}_{i=1,\ldots,d}$ of $A$, the conjugate state takes the form
\begin{equation}\langle  {\psi_T}\vert=\sum_{i_1\cdots i_N=1}^d\Tr[T(e_{i_1})\cdots T(e_{i_N})]\langle{i_1\cdots i_N}\vert\end{equation}

There may be many different ways to represent a given state in $A^{\otimes N}$ in an MPS form. Even the dimension of the virtual space $V$ is not uniquely defined. In general, it is not immediate to read off the properties of the state $\psi_T$ from the tensor $T$.

\begin{figure}[h]
\centering

\begin{tikzpicture}

\node at (0,0) {\ldots};

\draw[line width=2pt] (0.5,0)--node[above] {$\mu$} (2,0) circle(2pt)--node[above] {$\nu$} (4,0) circle (2pt)--node[above] {$\rho$} (6,0) circle(2pt)--node[above] {$\sigma$} (7.5,0);
\node at (8,0) {\ldots};

\draw (2,0)--(2,2) node[pos =0.5, left]{$T^{i}_{\ \mu \nu}$} node[right] {$i$}; \draw (4,0)--(4,2) node[pos =0.5, left] {$T^j_{\ \nu \rho}$} node[right] {$j$}; \draw (6,0)--(6,2) node[pos =0.5, left] {$T^k_{\ \rho \sigma}$} node[right] {$k$};

\end{tikzpicture}

\caption{An MPS represented as a tensor network} \label{fig:f1}

\end{figure}
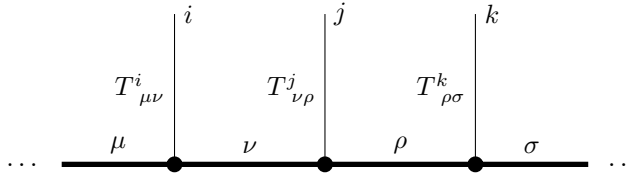

For the tensor $T$, one can construct a LCP Hamiltonian $H_T$, called the \emph{parent Hamiltonian}\footnote{There is a more general notion of a parent Hamiltonian where $h$ is any operator with this kernel; however, we will always take $h$ to be the projector.} of $\ket{\psi_T}$, which has $\ket{\psi_T}$ as a ground state. It is given as a sum of $2$-site terms $h_{s,s+1}$ that project onto the orthogonal complement of
$\ker h=(\cP\otimes\cP)(V\otimes\ket{\omega}\otimes V^*)$. Explicitly,\begin{align}\label{parent}&H_T=\sum_s h_{s,s+1}\hspace{5mm}\text{ where }\nonumber\\&h_{s,s+1}=\mathds{1}-(\cP_s\otimes\cP_{s+1})\delta(\cP_s^+\oplus\cP_{s+1}^+)\end{align}where $\delta$ is the projector onto $(V_s\otimes\ket{\omega}\otimes V_{s+1}^*)$ and $\cP_s^+:=(T_s
\cP_s)^{-1}T_s$ is a left inverse of $\cP_s$. The local projectors $h_{s,s+1}$ commute, so $H_T$ is gapped. $\ket{\psi_T}$ is annihilated by $h_{s,s+1}$, $\forall s$ and therefore also by $H_T$.

In general, $H_T$ has other ground states. Consider a state of the form\begin{align}\label{X-state}\ket{\psi_T^X}=&\left(\cP_1\otimes\cP_2\otimes \cdots\otimes\cP_N\right)\nonumber\\&\hspace{5mm}\left(\ket{\omega}_{12}\otimes\ket{\omega}_{23}\otimes\cdots\otimes\ket{\omega^X}_{N1}\right)\end{align}for some virtual state \begin{equation}\ket{\omega^X}=\sum_{i=1}^DX_{ij}\ket{i}\otimes\ket{j}\in V^*\otimes V\end{equation}where $X$ is a matrix that commutes with $T(a)$ for all $a\in A$. Note that $\ket{\omega^\mathds{1}}=\ket{\omega}$ and so $\ket{\psi_T^\mathds{1}}=\ket{\psi_T}$. The states \eqref{X-state} are clearly annihilated by $h_{s,s+1}$ for $s\ne N$. To see that they are annihilated by $h_{N1}$, note that tensor $T(e_i)XT(e_j)$ is expressible as a linear combination of tensors $T(e_i)T(e_j)$ if and only if $X$ commutes with every $T(e_i)$. The conjugate states have wavefunctions\begin{equation}\label{GS}\langle{\psi_T^X}\vert =\sum \Tr[X^\dagger T(e_{i_1})\cdots T(e_{i_n})]\langle{i_1\cdots i_N}\vert\end{equation}
We will refer to these states as \emph{generalized MPS}.

It turns out that all ground states of $H_T$ can be written as generalized MPS. One can always take $T$ to be an isometry with respect to some inner product on $A$ and the standard inner product
\begin{equation}\langle M|N\rangle =\Tr[M^\dagger N]\quad M,N\in\End(V)
\end{equation}on $\End(V)$. For an orthogonal basis $\{e_i\}$ of $A$, $\Tr[T(e_i)^\dagger T(e_j)]=\delta_{ij}$. Consider the case $N=1$. An arbitrary state
\begin{equation}\bra{\psi}=\sum_i a_i\bra{i}
\end{equation}can be written in generalized MPS form \eqref{GS} if one takes
\begin{equation}X=\sum_ja_jT(e_j)^\dagger\end{equation}Thus generalized MPS with commuting $X$ are the only ground states. Neither the number of generalized MPS nor the number of ground states depends on $N$; thus, the argument extends to all $N$.

Suppose the data $(A_1,V_1,T_1)$ and $(A_2,V_2,T_2)$ define two MPS systems with parent Hamiltonians $H_1$ and $H_2$. Consider the composite system $(A_1\otimes A_2,V_1\otimes V_2,T_1\otimes T_2).$ It has $\cP=\cP_1\otimes\cP_2$ and $\delta=\delta_1\otimes\delta_2$. Then\begin{eqnarray}h_{A\otimes B}&=&\mathds{1}_{A_1\otimes A_2}-\cP^2\delta\cP^{+2}_{A_1\otimes A_2}\nonumber\\&=&\mathds{1}_{A_1}\otimes\mathds{1}_{A_2}-\cP^2\delta\cP^{+2}_{A_1}\otimes \cP^2\delta\cP^{+2}_{A_2}\nonumber\\&=&(\mathds{1}_{A_1}-\cP^2\delta\cP^{+2}_{A_1})\otimes\mathds{1}_{A_2}+\mathds{1}_{A_1}\otimes(\mathds{1}_{A_2}-\cP^2\delta\cP^{+2}_{A_2})\nonumber\\&=&h_{A_1}\otimes\mathds{1}_{A_2}+\mathds{1}_{A_1}\otimes h_{A_2}\end{eqnarray}where the penultimate line follows from the fact that $\cP^2\delta\cP^{+2}$ is a projector. Therefore, the composite parent Hamiltonian is\begin{equation}\label{stackedham}H_{A\otimes B}=H_{A_1}\otimes\mathds{1}_{A_2}+\mathds{1}_{A_1}\otimes H_{A_2}.\end{equation}

\subsection{RG-fixed MPS and gapped phases}

Under real-space renormalization group (RG) flow,\cite{RG} adjacent pairs of sites are combined into blocks with physical space $A\otimes A$. The MPS form of the state is preserved, with the new MPS tensor being
\begin{equation}
T'(a\otimes b)=T(a)T(b),
\end{equation}
where on the r.h.s. the multiplication is matrix multiplication. We also define $\cP^\prime=T'^\dagger$. Though an RG step squares the dimension of the codomain of the MPS tensor, the rank is bounded above by $D^2$, and so the truncated physical space $\im(\cP^\prime)$ never grows beyond dimension $D^2$. 

An \emph{RG fixed MPS tensor} is an MPS tensor such that $\cP$ and $\cP^\prime$ have isomorphic images and are identical (up to this isomorphism) as maps. That is, there exists an injective map $\mu: A\ra A\otimes A$ such that
\begin{equation}
\mu\circ \cP=\cP^\prime.
\end{equation}
If we denote $m=\mu^\dagger$, this is equivalent to
\begin{equation}
T(m(a\otimes b))=T(a)T(b).
\end{equation}

Since $T$ was assumed to be injective, this equation completely determines $m$. Similarly, the fact that matrix multiplication is associative implies that $m:A\otimes A\ra A$ is an associative multiplication on $A$. The map $T:A\ra \End(V)$ then gives $V$ the structure of a module over $A$. Since $T$ is injective, this module is faithful (all nonzero elements of $A$ act nontrivially). The statement that $X$ commutes with $T$ in the ground state of the parent Hamiltonian is the statement that $X$ is a module endomorphism of $V$.

As previously stated, a state in $A^{\otimes N}$ may have multiple distinct MPS descriptions. One can always choose $T$ to have a certain \emph{standard form}\cite{MPS} -- regardless of whether it is RG fixed. When this is done, the matrices $T(a)$ are simultaneously block-diagonalized, for all $a\in A$. Moreover, if we denote by $T^{(\alpha)}$ the $\alpha^{\rm{th}}$ block, say of size $L_\alpha\times L_\alpha$, then the matrices $T^{(\alpha)}(e_i)$ span the space of $L_\alpha\times L_\alpha$ matrices. That is, $T^{(\alpha)}$ defines a surjective map from $A$ to the space of $L_\alpha\times L_\alpha$ matrices.

For an RG-fixed MPS tensor in its standard form, one can easily see that $A$ is a direct sum of matrix algebras. Indeed, each block $A^\alpha$ defines a surjective homomorphism $T^\alpha$ from $A$ to the algebra of $L_\alpha\times L_\alpha$ matrices, and if an element of $A$ is annihilated by all these homomorphisms, then it must vanish. Thus we get a decomposition
\begin{equation}\label{sumofmatrixalgebras}
A=\oplus_{\alpha} A^\alpha,
\end{equation}
where each $A^\alpha=(\ker T^\alpha)^\perp$ is isomorphic to a matrix algebra. We stress that some of these homomorphisms might be linearly dependent, so the number of summands may be smaller than the number of blocks in the standard form of $T$. An algebra of such a form is semisimple, that is, any module is a direct sum of irreducible modules. More specifically, any module over a matrix algebra of $L\times L$ matrices is a direct sum of several copies of the obvious $L$-dimensional module. This basic module is irreducible. If, for a particular $A^\alpha$, $T$ contains more than one copy of the irreducible  module, the corresponding blocks in the standard form of $T$ are not independent.

The ground-state degeneracy is simply related to the properties of the algebra $A$. Namely, the number of ground states is equal to the number of independent blocks in a standard-form MPS, or equivalently the number of summands in the decomposition (\ref{sumofmatrixalgebras}). Since the center of a matrix algebra consists of scalar matrices and thus is isomorphic to $\CC$, one can also say that the number of ground states is equal to the dimension of $\cA=Z(A)$.

Two gapped systems are said to be in the same phase if their Hamiltonians can be connected by a Local Unitary (LU) evolution, i.e. if they are related by conjugation with a finite-time evolution operator for a local time-dependent Hamiltonian.\cite{ChenGuWenzero} Clearly, the ground-state degeneracy is the same for all systems in a particular phase. In fact, for 1+1d gapped bosonic systems, it completely determines the phase.\cite{MPS,ChenGuWenone}

It is convenient to introduce an addition operation $\oplus$ on systems and phases. Given two 1+1d systems with local Hilbert spaces $A_1$ and $A_2$, we can form a new 1+1d system with the local Hilbert space $A_1\oplus A_2$. The Hamiltonian is taken to be the sum of the Hamiltonians of the two systems plus projectors which enforce the condition that neighboring ``spins'' are either both in the $A_1$ subspace or in the $A_2$ subspace. The ground state degeneracy is additive under this operation. A phase is called decomposable if it is a sum of two phases, otherwise it is called indecomposable. Clearly, it is sufficient to classify indecomposable phases.

It is easy to see that if $A$ decomposes as a sum of subalgebras, the corresponding phase is decomposable. Further, an indecomposable semisimple algebra $A$ is isomorphic to a matrix algebra. The corresponding ground state is unique. Moreover, while the parent Hamiltonians for different matrix algebras are different, they all correspond to the same phase,\cite{ChenGuWenone} i.e. are related by a Local Unitary evolution. Hence the phase is determined by the number of components in the decomposition \eqref{sumofmatrixalgebras}, or in other words, by $Z(A)$.

\section{Topological Quantum Field Theory}

We have seen above that an RG-fixed MPS state is associated with a finite-dimensional semisimple algebra $A$, and that the universality class of the corresponding phase depends only on the center of $A$. On the other hand, it is known since the work of Fukuma, Hosono, and Kawai \cite{FHK} that for any finite-dimensional semisimple algebra $A$ with an invariant scalar product one can construct a unitary 2D TQFT, and that the isomorphism class of the resulting TQFT depends only on the center of $A$. In this section we show that this is not a mere coincidence, and that the ground states of this TQFT can be naturally written in an MPS form, with an RG-fixed MPS tensor.

\subsection{State-sum construction of 2d TQFTs}

A (closed) 2D TQFT associates a space of states $\cA$ to an oriented circle, and a vector space $\cA^{\otimes n}$ to $n$ disjoint oriented circles. Further, suppose we are given an oriented bordism from $n$ circles to $l$ circles, i.e. a compact oriented 2d manifold $\Sigma$ whose boundary consists of $l$ circles oriented in the same way as $\Sigma$ and $n$ circles oriented in the opposite way. A 2d TQFT associates to $\Sigma$ a linear map from $\cA^{\otimes n}$ to $\cA^{\otimes l}$. This map is invariant under diffeomorphisms. Also, gluing bordisms taking care that orientations agree corresponds to composing linear maps. 

Let us mention some special cases. If $\Sigma$ is closed (i.e. has an empty boundary), then the 2D TQFT associates to it a linear map $\CC\ra\CC$, i.e. a complex number $Z_\Sigma$, called the partition function. If $\Sigma$ is a pair-of-pants bordism from two circles to one circle, the corresponding map $m: \cA\otimes \cA\ra \cA$ defines an associative, commutative product on $\cA$.  The cap bordism defines a symmetric trace function ${\rm Tr}:\cA\ra \CC$ such that the scalar product 
$\eta(a,b)={\rm Tr}(ab)$ is symmetric and non-degenerate. These data  make $\cA$ into a commutative Frobenius algebra.  It is known that a two-dimensional TQFT is completely determined by the commutative Frobenius algebra structure on $\cA$.\cite{Atiyah,MS,Abrams} The state-operator correspondence identifies $\cA$ with the algebra of local operators. This Frobenius algebra encodes the $2$- and $3$-point functions on the sphere, from which all other correlators, including the partition function, can be reconstructed. 

In 2d there is an essentially trivial family of unitary oriented TQFTs parameterized by a positive real number $\lambda$. The partition function of such a TQFT on a closed oriented 2d manifold $\Sigma$ is $\lambda^\chi(\Sigma)$, while the Hilbert space attached to a circle is one-dimensional. Such 2d TQFTS are called invertible, since the partition function is a nonzero number for any $\Sigma$. Since, by the Gauss-Bonnet theorem, $\chi(\Sigma)$ can be expressed as an integral of scalar curvature, tensoring a 2d TQFT by an invertible 2d TQFT is equivalent to redefining the TQFT action by a local counterterm which depends only on the background curvature. One usually disregards such counterterms. In what follows we will follow this practice and regard TQFTs related by tensoring with an invertible TQFT as equivalent.

Every unitary oriented 2d TQFT\footnote{More precisely, every equivalence class of unitary oriented 2d TQFTs, in the sense explained in the previous paragraph.} has an alternative construction called the state-sum construction,\cite{FHK} which is combinatorial and manifestly local. The input for this construction is a finite-dimensional semisimple algebra $A$, which is not necessarily commutative. To compute the linear maps associated to a  particular bordism $\Sigma$, one needs to choose a triangulation of $\Sigma$. Nevertheless, the result is independent of the choice of the triangulation. The connection between the not-necessarily commutative algebra $A$ and the commutative algebra $\cA$ is that $\cA$ is $Z(A)$, the center of $A$. From the perspective of open-closed TQFTs, $A$ is the algebra of states on the interval for a particular boundary condition. The scalar product on $\cA$ is also fixed by the structure of $A$. 

Let us describe the state-sum construction for the partition function $Z_\Sigma$ of a closed oriented 2D manifold $\Sigma$, following FHK.\cite{FHK} Fix a basis $e_i,$ $i\in S$, of $A$. We define the following tensors:\begin{equation}\label{gandC}
\eta_{ij}=\eta(e_i,e_j)=\Tr_A P_i P_j,\qquad C_{ijk}=\Tr_A P_i P_j P_k
\end{equation}
Here $P_i:A\ra A$ is the operator of multiplication by $e_i$.
The tensor $\eta_{ij}$ is symmetric and non-degenerate (if the algebra $A$ is semi-simple); the tensor $C_{ijk}$ is cyclically symmetric. We also denote by $\eta^{ij}$ the inverse to the tensor $\eta_{ij}$. Note also that $C_{ijk}$ is related to the structure constants $C^i{}_{jk}$ in this basis by
\begin{equation}
C^i{}_{jk}=\sum_l \eta^{il}C_{ljk}.
\end{equation}

Let $T(\Sigma)$ be a triangulation of $\Sigma$. A coloring of a 2-simplex $F$ of $T(\Sigma)$ is a choice of a basis vector $e_i$ for each 1-simplex $E\in \partial F$. A coloring of $T(\Sigma)$ is a coloring of all 2-simplices of $T(\Sigma)$. Note that each 1-simplex of $T(\Sigma)$ has two basis vectors attached to it, one from each 2-simplex that it bounds. The weight of a coloring is the product of $C_{ijk}$ over 2-simplices and $\eta^{ij}$ over 1-simplices, where the cyclic ordering of indices for each 2-simplex is determined by the orientation of $\Sigma$. The partition function is the sum of these weights over all colorings. 

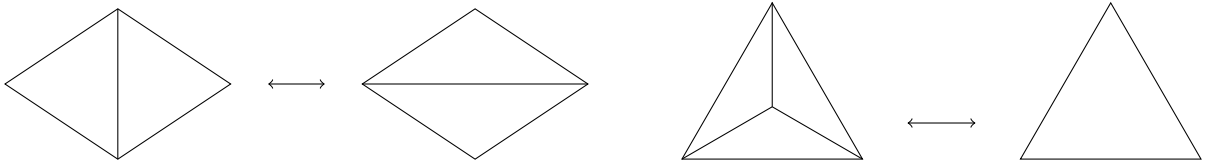
\begin{figure}
\centering
\begin{tikzpicture} [scale =0.5]

\draw (0,0)--(3,2)--(6,0)--(3,-2)--cycle; \draw (3,2)--(3,-2);

\draw [<->] (7,0)--(8.5,0);

\draw (9.5,0)--(12.5,2)--(15.5,0)--(12.5,-2)--cycle; \draw (9.5,0)--(15.5,0);

\end{tikzpicture} \hspace{1cm}
\begin{tikzpicture}[scale = 0.6]

\draw (0,0)--(2,2*1.732)--(4,0)--cycle;
\draw (0,0)--(2,2*0.58)--(4,0);
\draw (2,2*0.58)--(2,2*1.738);
\draw[<->] (5,0.8)--(6.5,0.8);

\begin{scope}[shift={(7.5,0)}]

\draw (0,0)--(2,2*1.732)--(4,0)--cycle;
\end{scope}

\end{tikzpicture}
\caption{The 2-2 and the 3-1 Pachner moves} \label{fig:f2} \end{figure}

Topological invariance of $Z_\Sigma$ can be shown as follows. It is known that any two triangulations of a smooth manifold are related by a finite sequence of local moves.\cite{Pachner} In two dimensions, there are two moves - the 2-2 move and the 3-1 move, depicted in Figure \ref{fig:f2} - which swap two or three faces of a tetrahedron with their complement. Invariance of the state-sum under the 2-2 ``fusion'' move reads\begin{equation}\label{2-2}C_{ij}{}^pC_{pk}{}^l=C_{jk}{}^pC_{ip}{}^l\end{equation}Similarly the 3-1 move reads\begin{equation}\label{3-1}C_{i}{}^{mn}C_{nl}{}^kC^{l}{}_{mj}=C_{ij}{}^k\end{equation}These axioms are satisfied by any finite-dimensional semisimple algebra $A$;\cite{FHK} therefore, the partition sum is a topological invariant\footnote{In two dimensions, there is no difference between topological and smooth manifolds.}.

\subsection{Open-closed 2d TQFT}

So far we have discussed what is known as closed 2D TQFTs. That is, the boundary circles were interpreted as spacelike hypersurfaces, and thus each spatial slice had an empty boundary. The notion of a TQFT can be extended to incorporate spatial boundaries; such theories are called open-closed TQFTs.
In such a theory a spatial slice is a compact oriented manifold, possibly with an nonempty boundary. That is, it is a finite collection of oriented intervals and circles. A bordism between such spatial slices is a smooth oriented \emph{surface with corners}: paracompact Hausdorff spaces for which each point has a neighborhood homeomorphic to an open subset of a half-plane. Surfaces with corners are homeomorphic, but typically not diffeomorphic, to smooth surfaces with a boundary. 

The corner points subdivide the boundary of the bordism into two parts: the initial and final spatial slices, and the rest. We will refer to the initial and final spatial slices as the cut boundary, while the rest will be referred to as the brane boundary. The cut boundary can be thought of as spacelike, while the brane boundary is timelike. Bordisms are composed along their cut boundary (hence the name), while on the brane boundary one needs to impose boundary conditions (known as D-branes in the string theory context, hence the name). More precisely, if $\cC$ is the set of boundary conditions, one needs to label each connected component of the brane boundary with an element of $\cC$. 

An open-closed 2d TQFT associates a vector space $V_{MM'}$ to every oriented interval with the endpoints labeled by $M,M'\in\cC$, and a vector space $\cA$ to every oriented circle. To a collection of thus labeled compact oriented 1D manifolds it attaches the tensor product of spaces $V_{MM'}$ and $\cA$. To every bordism with corners labeled in the way explained above, it attaches a linear map from a vector space of the `incoming'' cut boundary to the vector space of the ``outgoing'' cut boundary. Gluing bordisms along their cut boundaries corresponds to composing the linear maps. 

Just like in the case of a closed 2d TQFT, one can describe algebraically the data which are needed to construct a 2d open-closed TQFT. These axioms were discovered by Lazaroiu\cite{Lazaroiu}, and we also refer to Moore and Segal\cite{MS} for details. Suffice it to say that each space $V_{MM}$ is a (possibly noncommutative) Frobenius algebra, and each space $V_{MM'}$ is a left module over $V_{MM}$ and a right module over $V_{M'M'}$. That is, to every element $x\in V_{MM}$ one associates a linear operator $T^M(x): V_{MM'}\ra V_{MM'}$ so that composition of elements of $V_{MM}$ corresponds to the composition of linear operators: $T^M(x) T^M(x')=T^M(xx')$  (and similarly for $V_{M'M'}$).  Also, for every $M\in\cC$ there is a map $\iota^M: \cA\ra V_{MM}$ which is a homomorphism of Frobenius algebras. The dual map $\iota_M: V_{MM}\ra \cA$ is known as the generalized boundary-bulk map. In particular, if we act with $\iota_M$ on the identity element of the algebra $V_{MM}$, we get a distinguished element $\psi_M\in \cA$ called the boundary state corresponding to the boundary condition $M$. Geometrically, $\psi_M$ is the element of $\cA$ which the open-closed TQFT associates to an annulus whose interior circle is a brane boundary labeled by $M$, while the exterior circle is an outgoing cut boundary.

One may wonder if it is possible to reconstruct the open-closed TQFT from the closed TQFT. The answer turns out to be yes if $\cA$ is a semisimple, i.e. if every module over $\cA$ is a sum of irreducible modules.\cite{MS} \footnote{This might seem like a rather uninteresting case, since by the Wedderburn theorem every commutative semisimple algebra is isomorphic to a sum of several copies of $\CC$. But as explained below unitarity forces $\cA$ to be semisimple. Also, in the case of TQFTs with symmetries and fermionic TQFTs the classification of semisimple algebras is  more interesting.} Then $\cC$ is the set of finite-dimensional modules over $\cA$, and $V_{MM'}$ is the space of linear maps from the module $M$ to the module $M'$ commuting with the action of $\cA$ (i.e. $V_{MM'}$ is the space of module homomorphisms). Conversely, one can reconstruct the algebra $\cA$ from any ``sufficiently large'' brane $M\in\cC$: if we assume that the module $M$ is faithful (i.e. all nonzero elements of $\cA$ act nontrivially), then $\cA=Z(V_{MM})$.

\begin{figure*}
\centering
\begin{tikzpicture}[scale=0.8]

\node at (0.5,-1.732/2) {\ldots};
\draw [line width =2pt] (0,0)--(3,0) circle (3pt) node[above] {$\mu$}--  node[below] {$i$} (4,-1.732) circle (3pt) node[below] {$\rho$}--node[below] {$j$} (5,0) circle (3pt) node[above] {$\nu$}--(8,0);
\draw (1,0)--(2,-1.732)--(3,0);
\draw (5,0)--(6,-1.732)--(7,0);
\draw(1,-1.732)--(7,-1.732);
\node at (7.5,-1.732/2) {\ldots};

\begin{scope}[shift={(9,-0.8)}]
\draw[<->] (0,0)--(1.5,0);
\end{scope}

\begin{scope} [shift={(12,0)}]
\node at (-0.5,-1.732/2) {\ldots};
\draw[line width=2pt] (0,0)--(1,0) circle (3pt) node[above] {$\mu$}--node[above] {$k$} (3,0) circle (3pt) node[above] {$\nu$}--(4,0);
\draw (0,-1.732)--(1,0)--node [below] {$i$} (2,-1.732)--node [below] {$j$} (3,0)--(4,-1.732);
\draw (-0.5,-1.732)--(4.5,-1.732);
\node at (5,-1.732/2) {\ldots};
\end{scope}

\end{tikzpicture}

\caption{An elementary shelling representing $T^\mu_{\rho i} T^\rho_{\nu j}  =C^k_{ij} T^\mu_{\nu k}$ \eqref{b2-2}. The thick line is a physical boundary.}
\label{fig:f3}
\end{figure*}
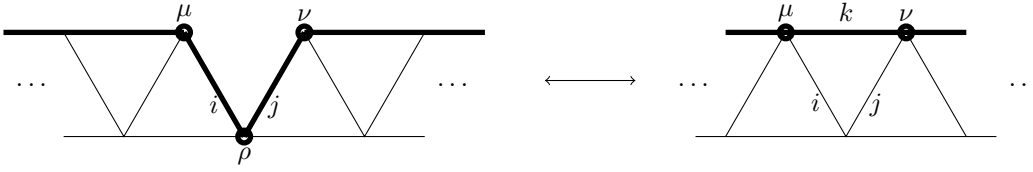

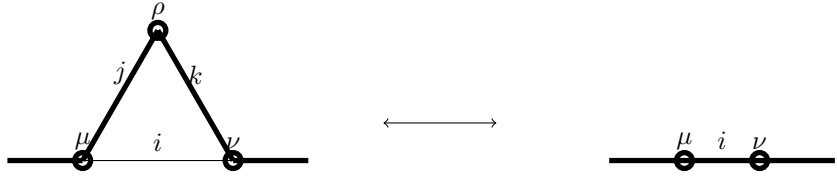
\begin{figure*}
\centering
\begin{tikzpicture}[scale=0.8]

\draw[line width=2pt] (1,0)--(2,0) circle (3pt) node[above]{$\mu$}--node[above] {$j$} (3, 1.732) circle (3pt) node[above]{$\rho$}--node[above] {$k$} (4,0) circle(3pt) node[above]{$\nu$}-- (5,0);
\draw (2,0)--node[above] {$i$} (4,0);

\draw[<->] (6,0.5)--(7.5,0.5);

\begin{scope}[shift = {(9,0)}]
\draw [line width=2pt] (0,0)--(1,0) circle (3pt) node[above] {$\mu$}-- node[above] {$i$} (2,0) circle (3pt) node[above] {$\nu$}--(3,0);
\end{scope}

\end{tikzpicture}
\caption{An elementary shelling representing $T^\mu_{\rho j}T^\rho_{\nu k} C^{jk}_i = T^\mu_{\nu i}$ \eqref{b3-1}.}
\label{fig:f4}
\end{figure*}

The state-sum construction generalizes to the open-closed case.\cite{LaudaSS} Let us describe it for a semisimple $\cA$, and assuming
that the bordism $\Sigma$ only has a brane boundary. Each connected
component of $\partial\Sigma$ is then labeled by a brane $M\in\cC$. We
pick a sufficiently large brane $M_0$ such that $\cA=Z(V_{M_0 M_0})$. Let
$A=V_{M_0M_0}$. We also choose a basis $f^M_\mu$, $\mu\in S_M$ in each
module $M$. Denote the matrix elements of the action of $A$ on $M$ by
$T_M^{\mu}{}_{\nu i}$. We choose a triangulation of $\Sigma$, which also
gives us a triangulation of each connected component of the boundary. 
2-simplices of $\Sigma$ are labeled as before. Label boundary 0-simplices
on any $M$-labeled boundary component by the basis vectors $f^M_\mu$. Thus
each boundary 1-simplex is labeled by a basis vector of $A$ and a pair of
basis vectors of a module. We assign a weight to each 2-simplex and each
interior 1-simplex before. We also assign a weight to each boundary 
1-simplex as follows. Suppose the boundary 1-simplex is labeled by 
$e_i\in A$ and $f^M_\mu,f^M_\nu\in M$. Then the weight of the boundary
1-simplex is $T_M^{\mu}{}_{\nu i}$. The total weight is the product of
weights of all 2-simplices and all 1-simplices (both interior and exterior).

Due to the introduction of brane boundaries, there are two more moves, called the 2-2 and 3-1 \emph{elementary shellings} and depicted in Figures \ref{fig:f3} and \ref{fig:f4}, that must be considered when demonstrating topological invariance. \cite{LaudaSS} They yield conditions
\begin{equation}\label{b2-2}T^\mu_{M\rho i} T^\rho_{M\nu j}  =C^k_{ij} T^\mu_{M\nu k}\end{equation}
and
\begin{equation}\label{b3-1}T^\mu_{M\rho j}T^\rho_{M\nu k} C^{jk}_i =T^\mu_{M\nu i}\end{equation}
respectively. The first one is the definition of a module, and the second one follows from the semisimplicity of $A$. Therefore the state-sum is a well-defined open-closed TQFT. Moreover, such structures are precisely those required to define a topologically invariant state-sum.

\subsection{Unitary TQFTs and semisimplicity}

The state-sum construction defines a perfectly good topological invariant for any finite-dimensional semisimple algebra $A$; however, if it is to model an actual physical system, its space of states must carry a Hilbert space structure, and linear maps corresponding to bordisms must be compatible in some sense with this structure. To be precise, for any oriented bordism $\Sigma$ whose source is a disjoint union of $n$ circles and whose target is a disjoint union of $l$ circles, let $-\Sigma$ denote its orientation-reversal. $-\Sigma$ has $l$ circles in its source and $n$ circles in its target. A 2d TQFT attaches to $\Sigma$ a linear map $\cA^{\otimes n}\ra \cA^{\otimes l}$, and to $-\Sigma$ a linear map $\cA^{\otimes l}\ra \cA^{\otimes n}$. A unitary structure on a 2d TQFT is a Hilbert space structure on $\cA$ such that the maps corresponding to $\Sigma$ and $-\Sigma$ are adjoint to each other. For an open-closed 2D TQFT, we require that the state-space assigned to each boundary-colored interval has a non-degenerate Hermitian metric, and that cobordisms with nonempty brane boundary also satisfy the Hermiticity condition. In particular, the product $m$ and coproduct $\mu$ are adjoints. It then follows from the Pachner moves that $\mu$ is an isometry. Likewise, the module structure $T$ is an isometry.

Let $\langle a,b\rangle$ denote the Hilbert space inner product of $a,b\in\cA$. Since $\cA$ also has a bilinear scalar product $\eta$, we can define an antilinear map 
\begin{equation}\label{CPT}
*: \cA\ra\cA, \quad a\mapsto a^*,
\end{equation}
such that $\langle a,b\rangle=\eta(a^*,b).$ It can be shown that this map is an involution (i.e. $a^{**}=a$) and an anti-automorphism (i.e. 
$(ab)^*=b^*a^*$).\cite{HQFT} This can also be expressed by saying that $\cA$ is a $*$-algebra. Conversely, one can show that any commutative Frobenius $*$-algebra such that the sesquilinear product $\eta(a^*,b)$ is positive-definite gives rise to a unitary 2d TQFT.\cite{HQFT}

A corollary of this result is that for a unitary 2d TQFT the algebra $\cA$ is semisimple. To see this, note first that any nonzero self-adjoint element $a$, $a=a^*$, cannot be nilpotent. Indeed, if $n$ is the smallest $n$ such that $a^n=0$, then $a^{2m}=0$, where $m=\lfloor(n+1)/2\rfloor$. Then $\left<a^m|a^m\right>=\left<1|a^{2m}|1\right>=0$, and therefore $a^m=0$. Since $n\leq m$, repeat with $n^\prime=m$ until $n=1$, i.e. $a=0$. Now we can use a result\cite{lifeQM} which says that a $*$-algebra with no nilpotent self-adjoint elements (apart from zero) is semisimple.

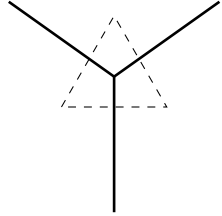
\begin{figure}[b]
\centering
\begin{tikzpicture}[scale=0.7]
\draw[dashed] (0,0)--(1,1.732)--(2,0)--cycle;

\draw[line width=1pt] (-1,2)--(1,1.732/3)--(3,2); \draw[line width=1pt] (1,1.732/3)--(1,-2);

\end{tikzpicture}
\caption{The Poincare dual of a triangle}\label{fig:f5}
\end{figure}

By the Artin-Wedderburn theorem, a finite-dimensional semisimple algebra over complex numbers is isomorphic to a sum of matrix algebras. Since $\cA$ is also commutative, this means that it is isomorphic to a sum of several copies of $\CC$. Frobenius and $*$-algebra structures exist and are unique up to isomorphism. This means that the only invariant of the 2d TQFT is the dimension of $\cA$, i.e. the ground-state degeneracy of the corresponding phase.

As discussed above, for a semisimple algebra $\cA$ boundary conditions correspond to finite-dimensional modules over $\cA$. It is easy to see that for the open-closed TQFT to be unitary, the algebra $V_{MM}$ must also have a Hilbert space structure such that\begin{equation}\label{unitarity}T(a)^\dagger=T(a^*).\end{equation}Such a structure always exists and is unique. Thus a boundary condition for a unitary 2d TQFT can be simply identified with a module over $\cA$. One can use any faithful module over $\cA$ as an input for the state-sum construction. 

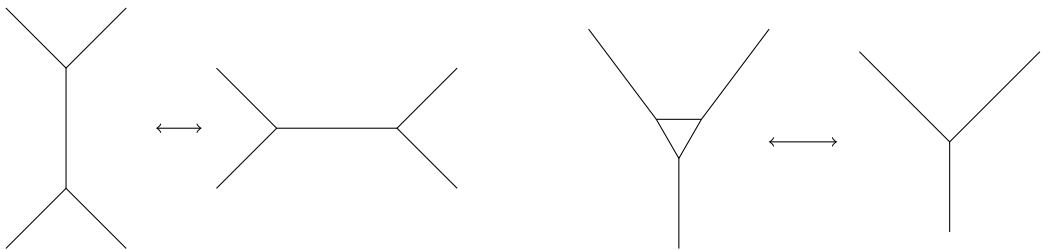
\begin{figure}
\centering
\begin{tikzpicture}[scale = 0.35]
\draw (0,2)--(2,0)--(4,2);
\draw (2,0)--(2,-4);
\draw (0, -6)--(2,-4)--(4,-6);

\begin{scope}[shift={(5,0)}]
\draw [<->] (0, -2)--(1.5,-2);
\end{scope}

\begin{scope}[shift={(7,-2)}]
\draw (0,2)--(2,0)--(0,-2);
\draw (2,0)--(6,0);
\draw (8,2)--(6,0)--(8,-2);
\end{scope}

\end{tikzpicture} \hspace{1.5cm}
\begin{tikzpicture}[scale=0.55]
\draw (0,1.4)--(1.5,0)--(2.5,0)--(4,1.4);
\draw (1.5,0)--(2,-1.732/2)--(2.5,0);
\draw (2,-1.732/2)--(2,-1.5-1.732/2);

\begin{scope}[shift={(4, -0.5)}]
\draw[<->] (0,0)--(1.5,0);
\end{scope}

\begin{scope}[shift={(6,-0.5)}]
\draw (0,2)--(2,0)--(4,2);
\draw (2,0)--(2,-2);
\end{scope}

\end{tikzpicture}
\caption{The dual 2-2 and 3-1 Pachner moves} \label{fig:f6} 
\end{figure}

\subsection{State-sum construction of the space of states}

We have discussed above the state-sum construction of the partition function $Z(\Sigma)$ for an oriented 2d manifold $\Sigma$ without boundary (or more generally, with only brane boundary). More generally, one also needs to describe in similar terms the state space $\cA$ and a linear map $\cA^{\otimes n}\ra \cA^{\otimes l}$ for every  bordism $\Sigma$ whose source is a disjoint union of $n$ circles and target is a disjoint union of $l$ circles. That is, one needs to describe $Z(\Sigma)$ for the case when $\Sigma$ has nonempty cut boundary. 

Consider a bordism $\Sigma$ with a nonempty cut boundary. For simplicity let us assume that there is no brane boundary; the general case is a trivial generalization, but requires a more cumbersome notation. We choose a triangulation $\cT$ of $\Sigma$. It induces a triangulation of each boundary circle. We label the edges of 2-simplices with basis elements of $A$, as before. The only difference is that boundary 1-simplices have only one label rather than two. If we assign the weights to every 2-simplex and every internal 1-simplex as before and sum over the labelings of internal 1-simplices, we get a number $Z_\cT(\Sigma)$ which depends on the labelings of the boundary 1-simplices. Suppose some boundary circle is divided into $N$ intervals. Then a labeling by $e_{i_1},\ldots, e_{i_N}$ corresponds to a vector 
\begin{equation}
e_{i_1}\otimes \ldots\otimes e_{i_N}\in A^{\otimes N}.
\end{equation}
We can think of the number $Z_{\cT}(\Sigma)$ computed by the state-sum as a matrix element of a linear map\begin{equation}A^{\otimes N_1}\otimes \ldots \otimes A^{\otimes N_n}\longrightarrow A^{\otimes M_1}\otimes \ldots\otimes A^{\otimes M_l},\end{equation}where $N_1,\ldots,N_n$ denote the number of 1-simplices in the source circles, and $M_1,\ldots,M_l$ denote the number of 1-simplices in the target circles of $\Sigma$. It can be shown \cite{FHK} that the map $Z_{\cT}(\Sigma)$ does not depend on the triangulation of $\Sigma$, provided we fix the triangulation of the boundary circles.

$Z_\cT(\Sigma)$ is not yet the desired $Z(\Sigma)$ because it depends on the way the boundary circles are triangulated. To get rid of this dependence, we need to restrict this map to a certain subspace in each source factor $A^{\otimes N_i}$ and project to a certain subspace in each target factor $A^{\otimes M_j}$. Both tasks are accomplished by means of projectors $C_N: A^{\otimes N}\ra A^{\otimes N}$. The  projector $C_N$ is simply $Z_{\cT_N}(C)$, where $C$ is a cylinder and $\cT_N$ is any triangulation of $C$ such that both boundary circle are subdivided into $N$ intervals. The image of each $C_N$ is a certain subspace of $A^{\otimes N}$ isomorphic to $Z(A)$.\cite{FHK} Restricting $Z_\cT(\Sigma)$  to these subspaces and then projecting to the image of each $C_{M_j}$ gives us the desired map
\begin{equation}
Z(\Sigma): \cA^{\otimes n}\ra \cA^{\otimes l},
\end{equation}
where $\cA=Z(A)$.

\subsection{MPS from TQFT}

Let us consider the special case when $\Sigma$ is an annulus such that one of the circles is a cut boundary, while the other one is a brane boundary corresponding to an $A$-module $M$. Let $T(a)\in\Hom(M,M)$ represent an action of $a\in A$ in this module. For definiteness, we choose the cut boundary to be the source of $\Sigma$, while the target is empty. Thus $Z(\Sigma)$ is a linear map $\cA\ra \CC$. It is the dual of the boundary state corresponding to the module $M$.

Let us now pick a triangulation of the annulus such that the cut boundary is divided into $N$ intervals. Then $Z_\cT(\Sigma)$ is a linear map $A^{\otimes N}\ra\CC$ which depends only on $\cT$ and $N$. We claim that this map is the dual of the MPS state with the dual MPS tensor given by $T: A\ra \Hom(M,M)$.

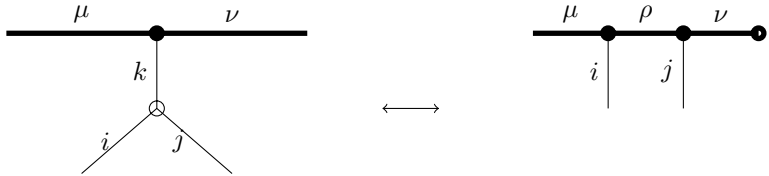
\begin{figure*}
\centering
\begin{tikzpicture}

\draw[line width=2pt] (0,0)--node [above] {$\mu$} (2,0) circle (2pt)--node[above] {$\nu$} (4,0);

\draw (2,0)--node[left] {$k$} (2,-1) circle (0.1)--node[left] {$i$} (1,-1-1.732/2);
\draw (2,-1)--node[left] {$j$} (3,-1-1.732/2);

\begin{scope}[shift={(5,-1)}]
\draw[<->] (0,0)--(0.75,0);
\end{scope}

\begin{scope}[shift={(7,0)}]
\draw[line width =2pt](0,0)--node[above]{$\mu$} (1,0) circle (2pt)--node[above] {$\rho$} (2,0) circle(2pt) --node[above]{$\nu$} (3,0) ;
\draw (1,0)--node[left] {$i$} (1,-1); \draw(2,0)--node[left] {$j$} (2,-1);
\end{scope}

\end{tikzpicture}
\caption{The dual shelling of \eqref{b2-2}. A filled dot represents $T$, while an empty dot represents $C$.}
\label{fig;f7}
\end{figure*}

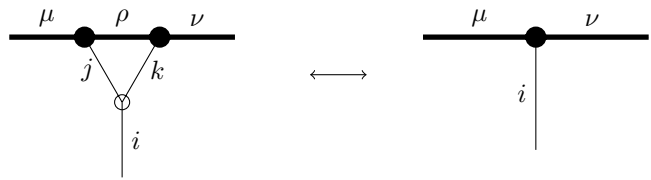
\begin{figure*}
\centering
\begin{tikzpicture}
\draw [line width=2pt,fill] (0,0)--node[above] {$\mu$} (1,0) circle (3pt) -- node[above] {$\rho$} (2,0) circle (3pt) -- node[above] {$\nu$} (3,0);

\draw (1,0)--node[left] {$j$} (1.5,-1.732/2) circle (0.1)--node[right] {$k$} (2,0);

\draw(1.5, -1.732/2)--node [right] {$i$} (1.5,-1.732/2 -1);

\draw[<->] (4,-0.5)--(4.75,-0.5);

\begin{scope}[shift={(5.5,0)}]
\draw [line width=2pt,fill] (0,0)--node[above] {$\mu$} (1.5,0) circle (3pt) -- node[above] {$\nu$} (3,0);
\draw (1.5,0)--node[left] {$i$}(1.5,-1.5);

\end{scope}

\end{tikzpicture}

\caption{The dual shelling of \eqref{b3-1}, representing $T^{j \ }_{\mu \rho} T^{k \  }_{\rho \nu} C_{ijk} = T^{i \ }_{\mu \nu}$ } \label{fig:f8}
\end{figure*}

To see this, it is convenient to reformulate the state-sum on the Poincare dual complex. This complex is built from the triangulation $\cT(\Sigma)$ by replacing $k$-cells with $(2-k)$-cells, as in Figure \ref{fig:f5}. The dual of a triangulation is not a simplicial complex but a more general cell complex; since we will only be interested in the edges and vertices of this dual complex, we will refer to it as a \emph{skeleton} for $\Sigma$. The Pachner moves are the same for skeleton as for triangulations, see Figure \ref{fig:f6}. Recall that for a unitary TQFT, one can choose $\eta_{ij}=\delta_{ij}$, so that indices may be freely raised and lowered; nonetheless, keeping track of index positions now will pay off later when we generalize to equivariant theories. Choose a direction for each edge; the state-sum does not depend on this choice. Choose these directions so that all edges on incoming boundaries are incoming and all edges on outgoing boundaries are outgoing. To define a state-sum on a skeleton, label its non-boundary edges with elements $e_i$ and assign structure coefficients $C$ to each non-boundary vertex according to orientation and using lower indices for incoming arrows and upper for outgoing. With these conventions, the Pachner moves algebrize to \eqref{2-2} and \eqref{3-1} as before. To incorporate brane boundaries, color brane boundary edges by elements $v_\mu$ and attach the module tensor $T$ to each boundary vertex. The boundary moves recover \eqref{b2-2} and \eqref{b3-1}. The dual state-sum is naturally a tensor network: it defines a circuit between the incoming and outgoing legs. Note that the ``virtual'' module indices are all contracted, so these legs are physical.

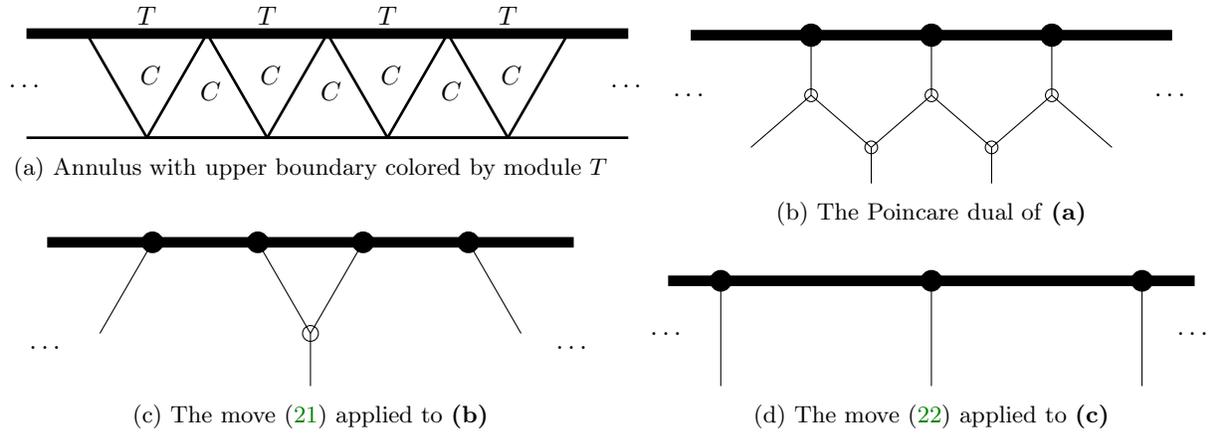
\begin{figure*}
\centering



\begin{subfigure}[b]{0.5\linewidth}
\centering

\begin{tikzpicture}[scale=0.75]


\draw[line width=4pt] (0,0)--(10,0);


\draw (1,0) -- (3,0) node[pos=0.5, above] {$T$}; 

\draw (1,0) -- (3,0) node[pos=1.5, above] {$T$}; 

\draw (1,0) -- (3,0) node[pos=2.5, above] {$T$}; 

\draw (1,0) -- (3,0) node[pos=3.5, above] {$T$}; 

\coordinate [label={above right:$C$}] (C) at (1.732,-1); \coordinate [label={above right:$C$}] (C) at (2.732,-1.27);  \coordinate [label={above right:$C$}] (C) at (3.732,-1); \coordinate [label={above right:$C$}] (C) at (4.732,-1.27); \coordinate [label={above right:$C$}] (C) at (5.732,-1); \coordinate [label={above right:$C$}] (C) at (6.732,-1.27);
\coordinate [label={above right:$C$}] (C) at (7.732,-1);


\draw[line width=1pt] (1,0) -- (2,-1.732) {}; 

\draw[line width=1pt] (2,-1.732) -- (3,0) -- (4,-1.732) -- cycle {};

\begin{scope}[shift={(2,0)}]

        \draw[line width=1pt] (2,-1.732) -- (3,0) -- (4,-1.732) -- cycle {};

\end{scope}

\begin{scope}[shift={(4,0)}]

        \draw[line width=1pt] (2,-1.732) -- (3,0) -- (4,-1.732) -- cycle {};

\end{scope}

\draw[line width=1pt] (9,0) -- (8,-1.732) {}; 

\draw[line width=1pt] (0,-1.732) -- (10, -1.732) {};

\node at (-0, -1.732/2) {\ldots};
\node at (10,-1.732/2) {\ldots};

\end{tikzpicture}

\caption{Annulus with upper boundary colored by module $T$}\label{fig:f9a}
\end{subfigure}%
\begin{subfigure}[t]{0.5\linewidth}
\centering

\begin{tikzpicture}[scale=0.75]

\draw[line width=4pt] (0,0)--(2,0) circle (3pt) --(4,0) circle (3pt) --(6,0) circle(3pt) --(8,0);
\draw (2,0)--(2,-1) circle (3pt)--(1,-1-1.732/2) ;
\draw (4,0)--(4,-1) circle(3pt);
\draw(2,-1)--(3,-1-1.732/2) circle(3pt)--(4,-1);
\draw (4,-1)--(5,-1-1.732/2) circle (3pt)--(6,-1)--(6,0);
\draw(6,-1) circle(3pt)--(7,-1-1.732/2);

\draw(3,-1-1.732/2)--(3,-1.6-1.732/2);
\draw (5,-1-1.732/2)--(5,-1.6-1.732/2);

\node at (8,-1) {\ldots};
\node at (0,-1) {\ldots};

\end{tikzpicture}

\caption{The Poincare dual of \textbf{(a)}}
\end{subfigure}%

\begin{subfigure}[b]{0.5\linewidth}
\centering

\begin{tikzpicture}[scale=0.66]

\draw [line width =4pt] (0,0)--(2,0) circle (3pt) -- (4,0) circle (3pt) -- (6,0) circle (3pt) --(8,0) circle (3pt)--(10,0);
\draw (2,0)--(1,-1.732);
\draw (4,0)--(5, -1.732) circle (0.15)-- (6,0); \draw (5,-1.732)--(5,-1-1.732);
\draw(8,0)--(9,-1.732);

\node at (10,-2) {\ldots};
\node at (0,-2) {\ldots};
\end{tikzpicture}

\caption{The move \eqref{b2-2} applied to \textbf{(b)}}
\end{subfigure}%
\begin{subfigure}[b]{0.5\linewidth}
\centering

\begin{tikzpicture}[scale=0.66]

\draw [line width =4pt] (0,0)--(1,0) circle (3pt) -- (5,0) circle (3pt) -- (9,0) circle (3pt) --(10,0);

\draw(1,0)--(1,-2);
\draw (5,0)--(5,-2);
\draw(9,0)--(9,-2);

\node at (10,-1) {\ldots};
\node at (0,-1) {\ldots};

\end{tikzpicture}

\caption{The move \eqref{b3-1} applied to \textbf{(c)}}
\end{subfigure}

\vfill

\caption{The equivalence of the annulus to the tensor network representation of an MPS} \label{fig:f9}
\end{figure*}

Consider the triangulation, shown in Figure \ref{fig:f9a}, of the annulus with boundary condition $T$ on one of its boundary components. Its state-sum defines a state in the physical space $\cA^N$. We claim that this state is the fixed point MPS $\ket{\psi_T}$. The proof of this fact is straightforward: by Pachner invariance, the annulus and MPS tensor networks are equivalent, see Figure \ref{fig:f9}.

More generally, one can insert a local observable on the brane boundary of the annulus. Such a local observable is parameterized by $X\in\Hom(M,M)$ which commutes with $T(a)$ for all $a\in A$. The corresponding dual state is $\Tr[X^\dagger TT\cdots T]$, i.e. it is a generalized MPS state, with $A$ being the physical space.

Since the linear operators $T(a)$ satisfy $T(a)T(b)=T(ab)$, all these MPS states are RG-fixed MPS states. The RG-step is described by the algebra structure on $A$, $m: A\otimes A\ra A$. Moreover, the MPS is automatically in a standard form. The module $T:A\ra\End(V)$ is semisimple, so it has a decomposition into simple modules $T^{(\alpha)}:A\ra\End(V^{(\alpha)})$. The collection of spaces $\End(V^{(\alpha)})$ form a block-diagonal subspace of $\End(V)$. Since $V^{(\alpha)}$ is simple, $T^{(\alpha)}$ surjects onto the block $\End(V^{(\alpha)})$. Moreover, as we have seen, unitarity of the TQFT enforces that $T$ is an isometry.

The parent Hamiltonian of the MPS on an $N$-site closed chain has a TQFT interpretation as well: it is the linear map $C_N=Z_{\cT_N}(C):A^{\otimes N}\rightarrow A^{\otimes N}$ assigned to a triangulated cylinder $C$ whose boundary consists of two circles triangulated into $N$ intervals. As previously stated, $C_N$ projects onto a subspace $\cA=Z(A)\subset A^{\otimes N}$, precisely the space of ground states of the parent Hamiltonian. In the continuum TQFT, topological invariance requires that the cylinder is the identity; this is consistent with our already having projected to $\cA$ in defining the continuum state spaces.

We have seen that a unitary TQFT is completely determined by its space of states $\cA$ on a circle and that each finite-dimensional commutative algebra $\cA$ defines a unitary TQFT. Therefore, the classification of unitary TQFTs is quite simple: there is one for every positive integer $n$, in agreement with the MPS-based classification of gapped phases. \cite{ChenGuWenone,ChenGuWentwo,FidkowskiKitaev}

\section{Equivariant TQFT and Equivariant MPS}

In this section, we generalize the relation between 2D TQFT and MPS states to systems with a global symmetry $G$. We show that both $G$-equivariant TQFTs and $G$-equivariant RG-fixed MPS states are described by semisimple $G$-equivariant algebras. In particular, we show that invertible $G$-equivariant TQFTs correspond to short-range entangled phases with symmetry $G$, and that both are classified by $H^2(G,U(1))$. 

\subsection{$G$-equivariant Matrix Product States}

Let $G$ be a finite symmetry group acting on the physical space $A$ via a unitary representation $R$, $g\mapsto R(g)\in\End(A)$. A $G$-invariant MPS tensor is a map $\cP: U\otimes U^*\ra A$ equivariant in the following sense: 
\begin{equation}
R(g)\cP(X)=\cP\left(Q(g) X Q(g^{-1})\right),
\end{equation}
where the linear maps $Q(g)\in\End(U)$ form a projective representation of $G$. Let $T=\cP^\dagger$. In terms of $T$, the equivariance condition looks as follows:
\begin{equation}
T(R(g)a)=Q(g) T(a) Q(g)^{-1},
\end{equation}
for any $a\in A$ and any $g\in G$.
The dual MPS state corresponding to $T$ is
\begin{equation}\label{equivMPS}
\langle \psi_T\vert=\sum_{i_1,\ldots,i_N}\Tr_U[T(e_{i_1})\ldots T(e_{i_N})] \langle i_1\ldots i_N\vert
\end{equation}
It is easy to see that the state $\psi_T$ is $G$-invariant, thanks to the equivariance condition on $P$. More generally, let $X\in\End(U)$. Note that $\End(U)$ is a genuine (not projective) representation of $G$. Then the generalized MPS state $\Tr[X TT\ldots T]$ transforms in the same way as $X$.

\subsection{$G$-equivariant TQFT}

Roughly speaking, a definition of a $G$-equivariant TQFT is obtained from the definition of an ordinary TQFT by replacing oriented manifolds with oriented manifolds with principal $G$-bundles. This reflects the intuition that a model with a global non-anomalous symmetry $G$ can be coupled to a background $G$ gauge field. (For a finite group $G$, there is no difference between a $G$ gauge field and a principal $G$-bundle.) 

Some care is required regarding marked points and trivializations. Namely, each source and each target circle must be equipped with a marked point and a trivialization of the $G$-bundle at this point. This means that the holonomy of the gauge field around the circle is a well-defined element $g\in G$, rather than a conjugacy class. A $G$-equivariant TQFT associates a vector space $\cA_g$ to a circle with holonomy $g$. A generic $G$-equivariant bordism has more than one marked point, and the holonomies between marked points along chosen paths are well-defined elements of $G$ as well. Of course, these holonomies depend only on the homotopy classes of paths. For example, a $G$-equivariant cylinder bordism has two marked points (one for each boundary circle) and depends on two arbitrary elements of $G$. 
On the other hand, a $G$-equivariant torus, regarded as bordism with an empty source and empty target, has no marked points and depends on two commuting elements of $G$ defined up to an overall conjugation. 

One can describe a $G$-equivariant TQFT purely algebraically in terms of a $G$-crossed Frobenius algebra.\cite{HQFT,MS} This notion generalizes the commutative Frobenius algebra $\cA$ and encodes the linear maps $Z(\Sigma,\cP)$ in a fairly complicated way. 

We will use instead a state-sum construction of 2D equivariant TQFTs which is manifestly local. Its starting point is a finite-dimensional semisimple $G$-equivariant algebra $A$. This is an algebra with an action of $G$ that preserves the multiplication $m: A\otimes A\ra A$. That is, $G$ acts on $A$ via a linear representation $R(g)$, $g\in G$, such that\begin{equation}\label{gauge1}m(R(g)a\otimes R(g) b)=R(g) m(a\otimes b).\end{equation}This condition implies that the group action also preserves the scalar product $\eta$ defined in \eqref{gandC}:\begin{equation}\label{etaortho}
\eta(R(g)a,R(g)b)=\eta(a,b).
\end{equation}The condition \eqref{etaortho} says that $R(g)$ is orthogonal with respect to $\eta$. As a consequence, if $R(g)$ commutes with the anti-linear map \eqref{CPT}, it is unitary with respect to the Hilbert space inner product.

A large class of examples of $G$-equivariant algebras is obtained by taking $A=\End(U)$, where $U$ is a vector space, and $G$ acts on $U$ via a projective representation $Q(g)$. It is clear that this gives rise to a genuine action of $G$ on $\End(U)$ which preserves the usual matrix multiplication on $\End(U)$. Moreover, the standard Frobenius structure
\begin{equation}
\eta(a,b)={\rm Tr}(ab)
\end{equation}
is clearly $G$-invariant. 

A $G$-equivariant module over a $G$-equivariant algebra $A$ is a vector space $V$ with compatible actions of both $A$ and $G$. That is, for every $a\in A$ we have a linear map $T(a):V\ra V$ such that $T(a)T(a')=T(aa')$, and for every $g\in G$ we have an invertible linear map $\Q(g):V\ra V$ such that $\Q(g)\Q(g')= \Q(gg')$. The compatibility condition that they satisfy reads\begin{equation}\label{gauge2}
T(R(g)a)=\Q(g)T(a)\Q(g)^{-1}\end{equation}
If we take $A=\End(U)$, where $U$ is a projective representation of $G$ with a 2-cocycle $\omega\in H^2(G,U(1))$, then $U$ is not a $G$-equivariant module over $A$ unless $\omega$ vanishes. However, if $W$ is a projective representation of $G$ with a 2-cocycle $-\omega$, then $U\otimes W$ is a $G$-equivariant module.\footnote{In fact, the category of projective representations of $G$ with a 2-cocycle $-\omega$ is equivalent to the category of $G$-equivariant modules over $\End(U)$, and the equivalence sends a projective representation $W$ to $U\otimes W$.}

Equivariant TQFTs admit a lattice description as well. It is simplest to describe a Poincare dual formulation in the sense of Section 3.5; spaces in this formulation also have direct interpretations as tensor networks. A trivialized background gauge field is represented on a skeleton as a decoration of each oriented edge with an element $g\in G$. Flipping the orientation of the edge replaces $g$ with $g^{-1}$. We require that the field is flat: that the product of the group elements around the boundary of each face is the identity element.\footnote{In the triangulation picture, we require the product of all group elements corresponding to edges entering a particular vertex to be the identity element.} In a basis $e_i$, $i\in S$ of $A$, the weight of a coloring of the skeleton is the product of the structure constants $C^{ijk}$ over vertices (with the cyclic order given by the orientation) and a factor $\eta(R(g)e_i,e_j)=R(g)^k{}_i\eta_{jk}$ for each edge directed from $i$ to $j$ labeled by $g$. The partition sum is the sum of these weights over all colorings; we emphasize that the group labels represent a background gauge field and are not summed. To incorporate brane boundaries, choose a $G$-equivariant module $V$ over $A$. Fix a basis $f_\mu$ of $V$. For each brane boundary vertex, label its adjacent boundary edges each with a basis element, so that each boundary edge has a total of two labels. The weight of a skeleton with a brane boundary is a product of $C$'s and $R$'s as well as a module tensor $T$ for each brane boundary vertex and a matrix element $\Q(g)^{\mu}_{\ \nu}$ for each brane boundary edge.

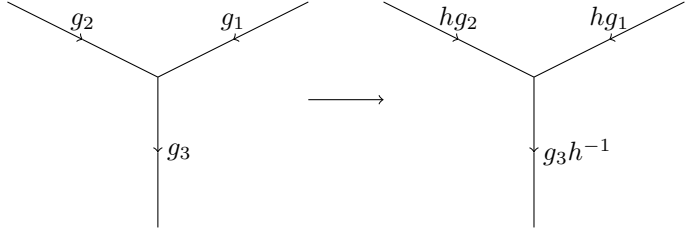
\begin{figure}
\centering
\begin{tikzpicture}[scale=0.9]
\draw[->] (-1,1)--(0,0.5) node[above] {$g_2$};
\draw (0,0.5)--(1,0);
\draw[->] (3,1)--(2,0.5) node[above] {$g_1$} ;
\draw (2,0.5)--(1,0);
\draw[->] (1,0)--(1,-1) node[right] {$g_3$};
\draw (1,-1)--(1,-2);

\draw[->] (3,-0.3)--(4,-0.3);

\begin{scope} [shift = {(5,0)}]
\draw[->] (-1,1)--(0,0.5) node[above] {$hg_2$};
\draw (0,0.5)--(1,0);
\draw[->] (3,1)--(2,0.5) node[above] {$hg_1$} ;
\draw (2,0.5)--(1,0);
\draw[->] (1,0)--(1,-1) node[right] {$g_3h^{-1}$};
\draw (1,-1)--(1,-2);
\end{scope}

\end{tikzpicture}
\caption{A gauge transformation at the vertex by $h$}
\label{fig:gauge}
\end{figure}

As before, topological invariance of the state-sum amounts to checking the conditions \eqref{2-2}, \eqref{3-1}, \eqref{b2-2}, and \eqref{b3-1}. These are satisfied by any finite-dimensional semisimple $A$. In order for the equivariant state-sum to constitute a well-defined equivariant TQFT, it must also be independent of the choice of trivialization of the background gauge field; in order words, it must be gauge invariant. A gauge transformation by $h\in G$ on a vertex acts by changing the decorations of the three edges whose boundary contains the vertex: incoming edges with $g$ become $hg$, outgoing $gh^{-1}$, as in Figure \ref{fig:gauge}. Invariance under a gauge transformation on a vertex in the interior is ensured by axioms \eqref{gauge1} and \eqref{etaortho} of a $G$-equivariant algebra. For vertices in the brane boundary, the analogous result follows from the $G$-equivariant module condition \eqref{gauge2}.\footnote{Here it is crucial that linear transformations $\Q(g)$ form an ordinary (i.e. not projective) representation of $G$.} Finally, invariance under simultaneously reversing an edge direction and inverting its group label is enforced by the axiom \eqref{etaortho}.

\subsection{$G$-equivariant semisimple algebras}

The classic Wedderburn theorem implies that every finite-dimensional semisimple algebra is a sum of matrix algebras. Let us discuss a generalization of this result to the $G$-equivariant case following Ostrik\cite{Ostrik} and Etingof\cite{Etingof}.

First, we can write every $G$-equivariant semisimple algebra as a sum of indecomposable ones, so it is sufficient to classify indecomposable $G$-equivariant semisimple algebras. A large class of examples is given by algebras of the form $\End(U)$, where $U$ is a projective representation of $G$. Another set of examples is obtained as follows: let $H\subset G$ be a subgroup. Consider the space of complex-valued functions on $G$ invariant with respect to left translations by $H$, i.e. $f(h^{-1}g)=f(g)$ for all $g\in G$ and all $h\in H$. The group $G$ acts on this space by right translations:
\begin{equation}
(R(g)f)(g')=f(g'g)
\end{equation}
Pointwise multiplication makes this space of functions into an associative algebra, and it is clear that the $G$-action commutes with the multiplication. This $G$-equivariant algebra is  indecomposable for any $H$.

The most general indecomposable $G$-equivariant semisimple algebra is a combination of these two constructions called the induced representation $\text{Ind}_H^G\End(U)$.\cite{Ostrik,Etingof} One picks a subgroup $H\subset G$ and a projective representation  $(U,Q)$ of $H$. Here $U$ is a vector space and $Q$ is a map $H\ra\End(U)$ defining a  projective action with a 2-cocycle $\omega\in H^2(H,U(1))$. Then one considers the space of functions on $G$ with values in $\End(U)$ which have the following transformation property under the left $H$ action:
\begin{equation}
f(h^{-1}g)=Q(h) f(g) Q(h)^{-1}
\end{equation}
It is easy to check that the right $G$ translations act on this space of functions. Pointwise multiplication makes this space into a $G$-equivariant algebra, and one can show that it is indecomposable. To summarize, indecomposable $G$-equivariant semisimple algebras are labeled by triples $(H,U,Q)$, where $H\subset G$ is a subgroup, and $(U,Q)$ is a projective representation of $H$. All these algebras are actually Frobenius algebras: the trace function $A\ra\CC$ is given by
\begin{equation}
\sum_{g\in G} {\rm Tr}_U f(g)
\end{equation}

A $G$-equivariant module over such an algebra $A$ is obtained as follows. Start with an $H$-equivariant module $(M,\Q)$ over $\End(U)$. Here $M$ is a module over $\End(U)$ and $\Q: H\ra\End(M)$ is a compatible  action of $H$ on $M$. As explained above, $M$ must have the form $U\otimes W$, where 
$W$ carries a projective action $S(h)$ of $H$ with a 2-cocycle $-\omega$.
Then consider functions on $G$ with values in $M$ which transform as follows under the left $H$-translations:
\begin{equation}
m(h^{-1}g)=(Q(h)\otimes S(h)) m(g),\quad m:G\ra U\otimes W.
\end{equation}
The group $G$ acts on this space by right translations, and it is easy to see that the pointwise action of $A=(H,U,Q)$ makes it into a $G$-equivariant module over $A$. One can show that any $G$-equivariant module over such an $A$ is a direct sum of modules of this sort. 

\subsection{$G$-equivariant MPS from $G$-equivariant TQFT}

It is sufficient to consider indecomposable TQFTs and $G$-equivariant algebras. Let us begin with the case $H=G$. Then the algebra $A=(G,U,Q)$ is isomorphic to the algebra $\End(U)$, and a $G$-equivariant module over it is simply a vector space $M$ with a $G$-equivariant action of $\End(U)$. In other words, $M=U\otimes W$, where $U$ carries a projective representation of $G$ with the 2-cocycle $\omega$, and $W$ carries a projective representation of $G$ with a 2-cocycle $-\omega$.

Consider an annulus whose outer boundary is labeled by a brane $M$ and whose inner boundary is a cut boundary. Let us triangulate both boundary circles into $N$ intervals. Let $g_{i,i+1}$ be the element of $G$ labeling the interval from the $(i+1)^{\rm th}$ to the $i^{\rm th}$ points on the boundary. We also assume that the holonomy of the gauge field between the points labeled by $1$ on the two boundary circles is trivial. We get the the following dual state:\begin{align}
\label{Gdec}\langle\psi_T\vert=\sum\Tr_{U\otimes W}[&T(e_{i_1})\Q(g_{1,2})\cdots\nonumber\\&\cdots T(e_{i_N}) \Q(g_{N,1})] \langle i_1 \cdots i_N \vert
.\end{align}
Note that although $T(e_i)$ is an operator on $U\otimes W$, it has the form $T(e_i)\otimes {\bf 1}_W$. Therefore, if $g_{i,i+1}=1$ for all $i$, the trace over $W$ gives an overall factor $\dim W$, and up to this factor we get the equivariant MPS (\ref{equivMPS}). Inserting an observable $X\in\End(U)$ on the brane boundary, we get a generalized equivariant MPS. The case when $X\in\End(U\otimes W)$ does not give anything new, since the trace over $V$ factors out.

The generalized equivariant MPS (cf. eq \ref{GS})
\begin{equation}
\langle \psi^X_T | = \sum \Tr[X^{\dagger} T(e_{i_1}) \cdots T(e_{i_n})] \langle i_1 \cdots i_n |
\end{equation}
may be charged under the action of $h\in G$:
\begin{widetext}
\begin{align}
R(h)^{\otimes N}\bra{\psi_T^X} = \sum \Tr[X^{\dagger} T(e_{i_1}) \cdots T(e_{i_n})] \langle (h^{-1} \cdot i_1) \cdots (h^{-1} \cdot i_n) | \nonumber \\
= \sum \Tr[X^{\dagger} T(h \cdot e_{i_1}) \cdots T(h \cdot e_{i_n})] \langle i_1 \cdots i_n | \nonumber \\ 
= \sum \Tr[Q(h^{-1}) X^{\dagger} Q(h) T(e_{i_1}) \cdots T(e_{i_n}) ] \langle i_1 \cdots i_n |
\end{align}
\end{widetext}

Let us now consider the case when $H$ is a proper subgroup of $G$ and $A=\text{Ind}_H^G\End(U)$, for some projective representation $U$ of $H$. If we choose right $H$-coset representatives $g_a$, $a\in H\backslash G$, and a basis $e_i$ in $\End(U)$, then a basis in $A$ is given by $e_i^a$. Similarly, if $f_\mu$ is a basis in an $H$-equivariant module $U\otimes W$, then a basis in the corresponding $G$-equivariant module $M$ is $f^a_\mu$. 

The action of $A$ on $M$ is diagonal as far as the $a$ index is concerned.  Therefore the dual state corresponding to a triangulated annulus with $g_{i,i+1}=1$ for all $i$  vanishes unless all $a$ indices are the same. Then
\begin{align}
\langle\psi_T\vert =\dim (W) &\sum_{a,i_1,\ldots,i_N} \Tr_U [T(e_{i_1})\cdots\nonumber\\&\hspace{5mm}\cdots T(e_{i_N})] \langle i_1 a\, i_2 a\,\cdots i_N a \vert.
\end{align}
This state has equal components along all $|H\backslash G|$ directions. We can get a state concentrated at a particular value of $a$ by inserting a suitable observable $X\in\End(M)$ on the brane boundary. Such an observable must commute with the action of $A$, so it must have the form $X^\mu_\nu { }^a_b= f(a)\delta^\mu_\nu\delta^a_b.$ Choosing the function $f(a)$ to be supported at a particular value of $a$ gives a generalized MPS state supported at this value of $a$.

The symmetry group $G$ acts transitively on $H\backslash G$. This suggests that we are dealing with a phase where the symmetry $G$ is spontaneously broken down to $H$, so that we get $|H\backslash G|$ sectors labeled by the index $a$.  To confirm this, consider the partition function of this TQFT on a  closed oriented 2-manifold $\Sigma$ with a trivial $G$-bundle. After we  choose a skeleton of $\Sigma$, we can represent this $G$-bundle by labeling every 1-simplex with the identity element of $G$. In addition, every 1-simplex is labeled by a pair of basis vectors of $A$. Since both the multiplication in the algebra $A$ and the scalar product are pointwise in $H\backslash G$, the partition function receives contributions only from those labelings where all $a$ labels are the same. Furthermore, turning on a gauge field which takes values in $H$ does not destroy this property. We conclude that the theory has superselection sectors labeled by elements of $H\backslash G$, and each sector has unbroken symmetry $H$.

\subsection{Twisted-sector states}

Now let us not assume that $g_{i,i+1}=1$, but instead allow the gauge field around the circle to have a nontrivial holonomy. Let us take $H=G$ first, i.e. the case of unbroken symmetry. Consider the MPS \eqref{Gdec}.  Applying a gauge transformations (by $g_{1,2}g_{2,3}\cdots g_{k-1,k}$ at vertex $k$) to the boundary vertices it can be written as
\begin{align} \langle\psi_{T,g}\vert=\sum &\Tr_{U\otimes W}[\Q(g)T(e_{i_1}) \cdots T(e_{i_N})]\nonumber\\&\hspace{5mm}\otimes_{k=1}^N R(g_{1,2}\cdots g_{k-1,k})^{i_k}{}_{j_k}\langle j_k \vert
\end{align}
where $g=g_{1,2}g_{2,3}\cdots g_{N,1}$ is the holonomy of the gauge field. This is LU equivalent to the state
\begin{equation} \langle\psi_{T,g}\vert=\sum \Tr_{U\otimes W}[\Q(g)T(e_{i_1}) \cdots T(e_{i_N})] \langle i_1 \cdots i_N \vert
\end{equation}
so we have effectively set $g_{i,i+1}=1$ for all $i\ne N$ and $g_{N,1}=g$. Note that $\Q = Q \otimes S$, so the trace factors into a product of a trace over $U$ and a trace over $W$. The latter gives us an overall factor, and we have
\begin{align}
 \langle\psi_{T,g}\vert =\Tr_W [S(g)] \sum \Tr_{U}[&Q(g)T(e_{i_1})  \cdots\nonumber\\&\cdots T(e_{i_N})] \langle i_1 \cdots i_N \vert.
\end{align}

This state transforms under $h \in G$ into
\begin{widetext}
\begin{align}
R(h)^{\otimes N}\bra{\psi_{T,g}} = (\Tr_W[S(g)]) \sum \Tr[Q(g)T(e_{i_1}) \cdots T(e_{i_n})] \langle (h^{-1} \cdot i_1) \cdots (h^{-1} \cdot i_n) | \nonumber \\
= (\Tr_W[S(g)])\sum \Tr[Q(h)^{ -1} Q(g) Q(h) T(e_{i_1}) \cdots T(e_{i_n})] \langle i_1 \cdots i_n | \nonumber \\
=(\Tr_W[S(g)]) \omega(g,h) \omega(h^{-1}, gh) \sum \Tr[ Q(h^{-1}gh)T(e_{i_1}) \cdots T(e_{i_n})] \langle i_1 \cdots i_n | \end{align}
\end{widetext}
Note that the $g$-twisted sector becomes the $hgh^{-1}$-twisted sector.

Now suppose $H$ is a proper subgroup of $G$. Since $T$ acts pointwise in the $a$ label, while $G$ acts on $a\in H\backslash G$ by right translations, the annulus state vanishes unless the holonomy around the circle is in $H$. This confirms once again that $H$ is the unbroken subgroup. Indeed, when the holonomy does not belong to the unbroken subgroup, there must be a domain wall somewhere on the circle. Its energy is nonzero in the thermodynamic limit, so the TQFT space of states must be zero-dimensional for holonomies not in $H$.

If $\cA_g$ denotes the space of states in the $g$-twisted sector, the space $\cA=\oplus_g\cA_g$ has an automorphism $\alpha_h:=R(h)^{\otimes N}$ for each $h\in G$ such that $\alpha_h(\cA_g)\subset\cA_{hgh^{-1}}$. $\cA$ is the $G$-graded vector space underlying the $G$-crossed Frobenius algebra that defines the associated $G$-equivariant TQFT.\cite{HQFT,MS}

\subsection{Morita equivalence}

We have seen that to any semisimple $G$-equivariant algebra one can associate a $G$-equivariant 2d TQFT. But different algebras may give rise to the same TQFT. In particular, we would like to argue that the TQFT corresponding to an indecomposable algebra $A=(H,U,Q)$, where $(U,Q)$ is a projective representation of $H$, depends only on the subgroup $H$ and the 2-cocycle $\omega$, but not on the specific choice of $(U,Q)$.

To show this, note first of all that the partition function vanishes if the holonomy does not lie in $H$ (this again follows from the fact that multiplication in the algebra $A$ is pointwise with respect to the $a$ index). Thus it is sufficient to consider oriented 2-manifolds with $H$-bundles. Further, if $U$ and $U'$ are projective representations of $H$ with the same 2-cocycle, then $U'=U\otimes W$, where $W$ is an ordinary representation of $H$. Thus we only need to show that the partition functions corresponding to algebras $(H,U,Q)$ and $(H,U\otimes W,Q\otimes S)$ are the same, where $S:H\ra\End(W)$ is a representation of $H$. But it is clear from the state sum construction that the two partition functions differ by a factor which is the partition function of two dimensional $H$-equivariant TQFT corresponding to the algebra $(H,W,S)$. 

We reduced the problem to showing that the $H$-equivariant TQFT constructed from the algebra $(H,W,S)$ is trivial when $(W,S)$ is an ordinary (not projective) representation of $H$. This is straightforward: the equation $S(h_1)\ldots S(h_n)=S(h_1\ldots h_n)$ and the flatness condition for the $H$ gauge field  imply that the partition function is independent of the $H$-bundle, and for the  trivial $H$-bundle the partition function is the same as for the trivial TQFT with $A=\CC$.

From the mathematical viewpoint, $G$-equivariant algebras with the same $H$ and $\omega$ are {\it Morita-equivalent}.\footnote{More accurately, algebras with the same $H$ and $\omega$, \emph{up to conjugation in} $G$, are Morita-equivalent. In physical contexts, however, it is typical to keep track of the embedding of the unbroken symmetry $H$ in the full symmetry group $G$. Therefore, the classification of physical gapped $G$-symmetric phases is slightly more refined than that of Morita classes.}\cite{Ostrik} Thus we have shown that Morita-equivalent algebras lead to identical $G$-equivariant TQFTs.\footnote{Strictly speaking, we only showed this for closed 2d TQFTs, but the argument easily extends to the open-closed case.}

\subsection{Stacking phases}

Consider two gapped systems built from algebras $A_1$ and $A_2$. Recall from Section \ref{sec:mps} that the stacked system \eqref{stackedham} is built from the tensor product algebra $A_1\otimes A_2$. Although we have not discussed parent Hamiltonians of $G$-equivariant MPS, an analogous stacking operation can be defined for $G$-symmetric gapped phases by way of the connection to TQFT. Now suppose $A_1$ and $A_2$ are $G$-equivariant algebras. It is clear from the $G$-equivariant state sum construction that the partition functions for the algebra $A_1\otimes A_2$ are products of those for $A_1$ and $A_2$ and that the Hilbert spaces are tensor products. Thus the MPS ground states, which determine a phase and which are realized in TQFT, stack like the tensor product of $G$-equivariant algebras.

It is a tedious but straightforward exercise to check that the result of stacking the phase labeled by subgroup-cocycle pair $(H,\omega)$ with the phase $(K,\rho)$ is the phase\begin{equation}\label{HK}(H\cap K,\omega|_{H\cap K}+\rho|_{H\cap K})^{\oplus[G:HK]}\end{equation}where $\omega|_{H\cap K}$ denotes the restriction of $\omega$ to the intersection subgroup $H\cap K$ and $[G:HK]$ denotes the index of the subgroup $HK$ in $G$, assuming $H$ and $K$ are normal in $G$.

Let us consider a simple example: take $G=\ZZ_2\times\ZZ_2=\langle a,b\rangle$, where $a$ and $b$ are commuting elements of order $2$. For the subgroup $H=G$, there are two cohomology classes $\omega\in H^2(\ZZ_2\times\ZZ_2,U(1))$. Let $\omega_1$ denote the nontrivial class. For each of the other subgroups $H=\langle a\rangle$, $\langle b\rangle$, $\langle ab\rangle$, $1$, there is a unique cocycle. Thus the classification of $\ZZ_2\times\ZZ_2$-equivariant phases is like Figure \ref{fig:classif}.

\begin{figure}
\centering
\begin{tabular}{ c | c | c }
$(H,\omega)$ & type of phase & name \\
 \hline			
  $(\langle a,b\rangle,1)$ & trivial & 1 \\
  $(\langle a,b\rangle,\omega_1)$ & symmetry-protected & $\omega$ \\
  $(\langle a\rangle,1)$ & broken symmetry & A \\
  $(\langle b\rangle,1)$ & broken symmetry & B \\
  $(\langle ab\rangle,1)$ & broken symmetry & C \\
  $(1,1)$ & broken symmetry & 0 \\
  \hline
\end{tabular}
\caption{Indecomposable phase classification for the $G=\ZZ_2\times\ZZ_2$}
\label{fig:classif}
\end{figure}

According to \eqref{HK}, the stacking rules are

\begin{widetext}
\begin{align}1\otimes 1=1,\quad 1\otimes\omega=\omega,\quad 1\otimes A=A,&\quad 1\otimes B=B,\quad 1\otimes C=C,\quad 1\otimes 0=0\nonumber\\\omega\otimes\omega=1,\quad\omega\otimes A=A,\quad\omega\otimes B&=B,\quad\omega\otimes C=C,\quad\omega\otimes 0=0\nonumber\\A\otimes A=A^{\oplus 2},\quad B\otimes B=B^{\oplus 2},\quad C\otimes C=&C^{\oplus 2},\quad A\otimes B=0,\quad B\otimes C=0,\quad C\otimes A=0\nonumber\\A\otimes 0=0^{\oplus 2},\quad B\otimes 0=0^{\oplus 2},&\quad C\otimes 0=0^{\oplus 2},\quad 0\otimes 0=0^{\oplus 4}\nonumber\end{align}
\end{widetext}

\subsection{Symmetry Protected Topological Phases}

Finally, let us discuss the case of Short-Range Entangled (SRE) phases with symmetry $G$. According to one definition,\cite{Kitaevtalk} an SRE phase is one that is invertible under the aforementioned stacking operation. Such phases have a one-dimensional space of ground states for every $G$-bundle on a circle. 
Since the space of states of a decomposable TQFT on a circle with a trivial bundle has dimension greater than one, a TQFT corresponding to an SRE phase must be indecomposable. We showed that when $H$ is a subgroup of $G$, the space of states is zero-dimensional whenever the holonomy does not lie in $H$. Hence an equivariant TQFT built from an indecomposable $G$-equivariant algebra $(H,U,Q)$ cannot correspond to an SRE unless $H=G$.

These SRE phases are all Symmetry Protected Topological (SPT) phases - phases that are trivial if we ignore symmetry. A $G$-equivariant algebra of the form $\End(U)$, where $U$ is a projective representation of $G$, is simply a matrix algebra if we ignore the $G$ action. Hence the corresponding non-equivariant TQFT is trivial; the corresponding Hamiltonian is connected to the trivial one by a Local Unitary transformation. Hence SPT phases with symmetry $G$ are labeled by 2-cocycles $\omega\in H^2(G,U(1))$. This is a well-known result. \cite{ChenGuWenone,ChenGuWentwo,FidkowskiKitaev}

\section*{ACKNOWLEDGEMENTS}

A. K. would like to thank P. Etingof and V. Ostrik for discussions. A.T. is grateful to I.~Saberi and D.~Williamson for helpful conversations. While this paper was nearing completion, we learned that closely related results have been obtained by K. Shiozaki and S. Ryu. This paper was supported in part by the U.S. Department of Energy, Office of Science, Office of High Energy Physics, under Award Number DE-SC0011632.

\end{document}